\documentclass[12pt]{article}
\usepackage{amsfonts}
\usepackage{amsmath}

\usepackage[english]{babel}
\usepackage[latin1]{inputenc}
\usepackage[T1]{fontenc}

\newtheorem{theorem}{Theorem}
\newtheorem{definition}{Definition}
\newtheorem{lemma}{Lemma}

\newtheorem{corollary}{Corollary}

\def\mathbb{\mathbf}

\begin{document}

\title{On the limit of superposition states}
\begin{center}
\centerline{LUIGI ACCARDI}

\textit{Centro Vito Volterra, Universit\`a  di Roma ''Tor Vergata'', \\
Roma I-00133, Italy}

first\_accardi@volterra.uniroma2.it



\centerline{ABDESSATAR SOUISSI}

\textit{$^1$ Department of Accounting, College of Business Management\\
Qassim University, Ar Rass, Saudi Arabia \\
$^2$ Preparatory Institute for Scientific and Technical Studies La Marsa,\\
 Carthage University, Tunisia}\\
second\_a.souaissi@qu.edu.sa, \quad  \ abdessattar.souissi@ipest.rnu.tn\\

\centerline{El GHETEB SOUEIDY}
\textit{Department of Mathematics,
Nouakchott University,\\ Nouakchott, Mauritania.}

third\_elkotobmedsalem@gmail.com
\end{center}

\maketitle

\tableofcontents

\begin{abstract}
In this paper, we study the structure  of a family of superposition states  on tensor algebras. The correlation functions of the considered states are described through a new kind of positive definite kernels valued in the dual of C$^\ast$-algebras, so-called Schur kernels. Mainly, we show the existence of the  limiting state of a net of  superposition states over an arbitrary  locally finite graph. Furthermore, we show that this limiting state enjoys a mixing property and an $\alpha$-mixing property  in the case of the multi-dimensional integer lattice $\mathbb{Z}^\nu$.\\

\end{abstract}

\section{Introduction}
It is well known that, in  statistical mechanics  and quantum field theory  superposition states allow the description of the subsystem relative to bounded regions  in terms of its degrees of freedom \cite{Emch}.\\
The superposition states in quantum theory are   convex combinations of  many states. In many important cases, a complex quantum  system is  identified with  a graph $V= (V, E)$ with vertex set $V$ and edge set $E$. To each bounded region $\Lambda\subset V$, we associated a finite dimensional Hilbert space $\mathcal{H}_\Lambda:= \bigotimes_{x\in \Lambda}\mathcal{H}_{x}$.
The local superposition states under consideration are states  on the algebras $\mathcal{B}(\mathcal{H}_\Lambda)$  indexed  by   bounded regions $\Lambda$ has the form
\begin{equation}\label{State_psi}
\psi_\Lambda (.) : = \langle \Psi_{\Lambda}, (.) \Psi_{\Lambda}\rangle
\end{equation}
for some state vector $\Psi_{\Lambda} :=\sum_{i\in I} \bigotimes_{x\in \Lambda} h_{x, i}\in\mathcal{H}_{\Lambda}$ where $I$ is a finite set and  $\{h_{x, i}\}_{i \in I}\subset \mathcal{H}_{x}$.\\
The present paper mainly deals with   the structure  of limiting state of a net of superposition states. Namely, we describe  the correlation  functions in terms of Schur multiplication. This  leads to  a class of positive definite Kernels with values in the dual of  C$^*$-algebras, so-called \textit{ Schur kernels}. \\
 Notice that positive definite kernels valued in C$^{*}$-algebras were studied in connection with operator-valued cocycles, see for instance \cite{J.Hoe08}. It turns out that, the Schur multiplication appears naturally in the explicit form of the states under consideration.These results might open new perspective to the  study of ergodic properties of this kind of states, such as entanglement  and mixing.\\
In \cite{[AcFi03-EMC]}, it was proven that  the states of the form  (\ref{State_psi})   have rich entanglement properties. Furthermore,  it was shown that they define a class of  quantum  Markov chains in the sense of \cite{[Ac74d-Camerino]}, \cite{[Ac74f-FAA]}.\\
The finite volume states (\ref{State_psi}) are states on type $I$ factors, with finite degree of freedom. In local quantum physics \cite{Haag96}, it was  proven that these states  play a key role in the description of quantum systems with an infinite degree of freedom, naturally appearing in quantum fields theory  and quantum statistical mechanics. Moreover, some relevance  of such kind of states in  quantum information theory,   quantum random walks and quantum hidden Markov  models were clarified ( see for instance \cite{[AcOh91-Compound channels]}, \cite{[AcMatsOhya051215-IDAQP]}, \cite{[AcMatsOhy08]},   \cite{[Kan19]}, \cite{[BeOs19]}).\\
It is crucial to know certain ergodic properties of  the limiting state  of the  superposition states (\ref{State_psi}), which may reveal some physical characteristics of the system. With this aim, we prove a mixing property for the inhomogeneous superposition states under consideration in the case of  the multidimensional integer lattice. Moreover, we exhibit a definition for $\alpha$-mixing property and show that the limit of superposition states is indeed $\alpha$-mixing.\\
A main problem in entanglement theory, is to find an easily applicable criteria to distinguish quantum states according to their degree of entanglement,  or at least to discriminate entangled  from non entangled states. The first step to attack this problem is to have a sufficiently detailed description  of the class of superposition states under consideration.\\
The paper is organized as follows.  We devote Section \ref{sec_schur_ker} to Schur kernels and their properties. In section \ref{sec_str_sup_state}, we describe the correlations functions of finite volume superposition states through a class of Schur kernels. Section \ref{sec-inf-vol-lim} is devoted to the existence (Theorem \ref{thm_limit_state}) of the  infinite volume-limiting state. Section \ref{sec-Homog-case} is devoted to the homogeneous case. Section \ref{Sec_mixing} is devoted to the mixed property for the inhomogeneous limiting state. This will be the aim of some futures works.
\section{Preliminaries}\label{sec_prel}
In this section we introduce some basic notions and notations.
Let $d\in \mathbb N$, the space of $d\times d$ matrices over the complex field $\mathbb C$ will be denoted $M_d(\mathbb{C})$.
Let us consider two C$^*$-algebras $\mathcal{A}$  and $\mathcal{B}$. Recall that a map $\Phi: \mathcal{A} \to \mathcal{B}$ is called \textit{positive} if it maps $\mathcal{A}^{+}$ into $\mathcal{B}^{+}$. A linear map $\Phi$ is called $k-$positive if the mapping $\Phi\otimes Id_k$ is positive from $M_k(\mathcal{A})\equiv \mathcal{A}\otimes M_k(\mathbb{C})$ into $M_k(\mathcal{B})\equiv \mathcal{B}\otimes M_k(\mathbb{C})$. If $\Phi$ is $k$--positive for each integer $k$ then it is called \textit{completely positive} (CP). It is well known that a linear map $\Phi$ is completely positive if and only if
\begin{equation}\label{CP-cac}
\sum_{i,j=1}^{k}b_{i}^{*}\Phi(a_{i}^{*}a_{j})b_{j}\ge 0; \quad   a_1, \cdots, a_k\in \mathcal{A},   b_1, \cdots, b_k\in \mathcal{B}
\end{equation}
for every $k\in \mathbb{N}$. The reader is referred to \cite{[Paulsen02]} for a detailed discussion about CP maps. The dual space of $\mathcal{B}$ will be denoted by $\mathcal{B}^{*}$ (see for instance \cite{[Sakai71]}).\\
Let $G= (V, E)$ be a simple graph. To each vertex $x\in V$ we associate a finite dimensional Hilbert space $\mathcal{H}_{x}$.
Let $\Lambda\subset V$ be a bounded region. To which we associate  a Hilbert space of observable $\mathcal{H}_\Lambda$ and the associated algebra of observable
\begin{equation}\label{df-B-Lambda-Bx}
\mathcal{B}_{\Lambda}
:=\mathcal{B}(\mathcal{H}_{\Lambda})
\equiv\bigotimes_{x\in \Lambda}\mathcal{B}(\mathcal{H}_{x})
=:\bigotimes_{x\in \Lambda}\mathcal{B}_{x}.
\end{equation}
Let $\mathcal{H}$ be a Hilbert space. The adjoint of a vector  $\xi\in \mathcal{H}$ is defined by
$\xi^{*} (\eta) = \langle \eta, \xi\rangle.$
The notation $\eta\xi^{*}$ refers to the rank-one operator over $\mathcal{H}$ defined by
$\eta\xi^{*}(\psi) = \langle \psi, \xi \rangle \eta.$\\
Throughout this paper, by $\mathrm{Tr}_{\mathcal{H}}$, we denote the trace on the algebra $\mathcal{B}(\mathcal{H})$ of all bounded operators over $\mathcal{H}$. Notice that for any orthonormal basis $\{e_{j}\}_{1\le j \le d} $ of the Hilbert space $\mathcal{H}$
$$
\mathrm{Tr}_{\mathcal{H}}(a)=\sum_{j=1}^{d}\langle ae_{j}, e_{j}\rangle, \quad \forall a\in  \mathcal{B}(\mathcal{H})
$$
where $d=\dim(  \mathcal{H})$.\\
The Schur multiplication of two matrices $A = (a_{i,j}), B = (b_{i,j}) \in M_d(\mathbb{C})$
is defined by
\begin{equation}\label{Schur_matrices}
A\diamond B_{i,j} = a_{i,j}b_{i,j}.
\end{equation}
The Schur algebra $(M_{d}(\mathbb{C}), \diamond)$ is abelian, with identity the matrix  $J$ with all entries are equal $1$.

\section{Schur kernels}\label{sec_schur_ker}
This section we introduce a new kind of positive definite kernels valued in dual of C$^*$-algebras. These kernels will be referred as   Schur kernels. We will  see later that these kernels play a crucial role in the  description of  correlation functions of the superposition states.
\begin{definition}\label{df-Schur-kern}{\rm
Let $I$ be a finite set and $\mathcal{B}$ a $C^*$--algebra. A map
$$
E: I\times I\to \mathcal{B}^*
$$
is called   \textbf{Schur kernel} associated to the pair $(I,\mathcal{B})$ if the family of linear functionals $\{E_{j,i} \}_{j,i\in I}$
enjoys the following property.\\
\textbf{Property S}: for any $n\in\mathbb{N}$ and any function
$$
b: h\in\{1, \dots,n\} \to b_{h}\in\mathcal{B}
$$
the complex valued kernel on $I\times \{1, \dots,n\}$ given by
\begin{equation}\label{Property-S}
K_{(j,h),(i,k)}:=
E_{j,i}\left(b_{h}^*b_{k}\right)
\quad;\quad  (j,h), (i,k)\in I\times \{1, \dots,n\}
\end{equation}
is positive definite.
}\end{definition}
\textbf{Remark}. The Schur kernels in the sense of Definition  \ref{df-Schur-kern} are valued in the dual of the C$^*$-algebra. Property (S) implies that for each fixed element $b\in \mathcal{B}$  the map
$$
K_b : (i,j)\in I^2  \to E_{i,j}(b^{*}b)
$$
is a complex valued positive definite kernel. And for any fixed $i\in I$ the map
$$
K_{i}: (h,k)\in \{1,2, \cdots, n\}^2 \to E_{i,i}(b_{h}^{*}b_k)
$$
is also a complex positive definite kernel.
\begin{theorem}\label{pos-mps-Schur-kern}{\rm
Let $I$ be a \textbf{finite set} $(d_I = |I|)$  and $\mathcal{B}$ a $C^*$--algebra. A map
$$
E: I\times I\to \mathcal{B}^*
$$
is a $(I,\mathcal{B})$--\textbf{Schur kernel} iff the map $\hat{E}$ defined by
\begin{equation}\label{df-Eji(b)}
\hat{E}: b\in \mathcal{B}\to
\hat{E}(b) :=(E_{i,j }(b))_{i,j\in I}\in M_{d_{I}}(\mathbb{C})
\end{equation}
is \textbf{completely positive}.
}\end{theorem}
\textbf{Proof}. \textbf{Necessity}. If $\hat{E}$ is completely positive then for any $a_1, \ a_2,\cdots,\ a_n \in M_{d_{I}}(\mathbb{C}) $ and $b_1, \ b_2,\cdots,\ b_n \in \mathcal{B} $
$$
\sum_{k,j=0}^{n} a_{j}^*\hat{E}(b_{j}^* b_{k})a_{k}\geq 0 \Leftrightarrow \ \sum_{k,j=0}^{n} a_{j}^*\left[E_{ \ \cdot \ , \ \cdot \ }(b_{j}^* b_{k})\right]a_{k}\geq 0
$$
where $\hat{E}(b_{j}^* b_{k})=\left[E_{ \ \cdot \ , \ \cdot \ }(b_{j}^* b_{k})\right]$.\\
By taking $a_{j}=(a_{k,l}^{(j)})_{k,l \in I}$, we have for each
$X =(x_1,\cdots,x_{d})\in M_{1\times d_{I}}(\mathbb{C})$ at
$$
X\left(\sum_{k,j=0}^{n} a_{j}^*\left[E_{ \ \cdot \ , \ \cdot \ }(b_{j}^* b_{k})\right]a_{k}\right)X^* \geq 0
$$
$$
\iff \sum_{k,j=0}^{n}X \left(\sum_{s=1}^{d}
\sum_{l=1}^{d}\overline{a}_{l,p}^{(j)}E_{l,s}(b_{j}^* b_{k})a_{s,q}^{(k)}\right)_{p,q \in I}X^*\geq 0
$$
\begin{equation}\label{CP}
\iff \sum_{k,j=0}^{n}
\left(\sum_{t,s,l,r=1}^{d}x_{t}\overline{a}_{l,t}^{(j)}E_{l,s}(b_{j}^*b_{k})a_{s,r}^{(k)}x^{*}_{r}\right)
\geq 0
\end{equation}
For  $X=E_{1}:= (1, 0, \cdots, 0)\in M_{1\times d_{I}}(\mathbb{C})$ one gets
$$
\sum_{t,s,l,r=1}^{d}x_{t}\overline{a}^{(j)}_{l,t}E_{l,s}(b_{j}^* b_{k})a_{s,r}^{(k)}x_{r}^{*}=\sum_{s,l=1}^{d}\overline{a}_{l,1}^{(j)}E_{l,s}(b_{j}^* b_{k})a_{s,1}^{(k)}.
$$
By taking $n=d$ and  denote $z_{j,l}=a_{l,1}^{(j)}$, then (\ref{CP}) becomes
$$
\sum_{(l,j),(s,k)\in I\times I}\overline{z}_{j,l}E_{l,s}(b_{j}^* b_{k})z_{k,s}\geq 0
$$
Since $a_{l,1}^{(j)}$ can be  arbitrary chosen, one concludes that $E$ is a Schur kernel.
\noindent \textbf{Sufficiency.} Conversely, if  $E$ is a  Schur kernel, let
$a_1, \ a_2,\cdots,\ a_n \in M_{d_{I}}(\mathbb{C}) $ and $b_1, \ b_2,\cdots,\ b_n \in \mathcal{B} $  be such that
$$
\sum_{k,j=0}^{n} a_{j}^*\hat{E}(b_{j}^* b_{k})a_{k}= \ \sum_{k,j=0}^{n} a_{j}^*\left[E_{ \ \cdot \ , \ \cdot \ }(b_{j}^* b_{k})\right]a_{k}
$$
Let $X =(x_1,\cdots,x_d)\in M_{1\times d_{I}}(\mathbb{C})$, then
$$
X\left(\sum_{k,j=0}^{n} a_{j}^*\left[E_{i,m}(b_{j}^* b_{k})\right]_{i,m}a_{k}\right)X^*
=
\sum_{k,j=0}^{n}\left(\sum_{s,l=1}^{d}\left(\sum_{t=1}^{d}x_{t}\overline{a}_{l,t}^{(j)}\right)E_{l,s}(b_{j}^* b_{k})\left(\sum_{r=1}^{d}a_{s,r}^{(k)}\overline{x}_{r}\right)\right)
$$
$$
=\sum_{k,j=0}^{n}\left(\sum_{s,l=1}^{d}\overline{z}_{j,l}E_{l,s}(b_{j}^* b_{k})z_{k,s}\right)=\sum_{(j,l),(s,k)\in I\times I}\overline{z}_{j,l}E_{l,s}(b_{j}^* b_{k})z_{k,s}
$$
$$
=\sum_{(j,l),(s,k)\in I\times I}\overline{z}_{j,l}K_{(j,l),(k,s)}z_{k,s}\geq 0
$$
where $z_{k,s}=\sum_{r=1}^{d}a_{s,r}^{(k)}\overline{x}_{r}$. $\square$
\begin{definition}\label{tens-prod--Schur-kerns}{\rm
Let $I$ be a set, $\mathcal{B}$, $\mathcal{C}$ two $C^*$--algebras and
$E_{\mathcal{B};  \ \cdot \ }$, $E_{\mathcal{C};  \ \cdot \ }$ two Schur kernels
respectively associated to the pairs $(I,\mathcal{B})$, $(I,\mathcal{C})$.
The \textbf{tensor product of the kernels} $E_{\mathcal{B}; \ \cdot \ }$ and
$E_{\mathcal{C}; \ \cdot \ }$, denoted
$E_{\mathcal{B}; \ \cdot \ }\otimes E_{\mathcal{C}; \ \cdot \ }$, is defined by
$$
(j,i)\in I\times I\mapsto E_{\mathcal{B}; j,i}\otimes E_{\mathcal{C}; j,i}\in
\mathcal{B}^* \otimes\mathcal{C}^* \equiv (\mathcal{B}\otimes\mathcal{C})^*
$$
where $E_{\mathcal{B}; j,i}\otimes E_{\mathcal{C}; j,i}$ is the usual tensor product of duals of $C^*$--algebras.
}\end{definition}
\textbf{Remark}. Definition \ref{tens-prod--Schur-kerns}, as well as Theorem \ref{tens-prod--Schur-kerns-is-SK} below, do not depend on the choice of the $C^*$--cross
norm used to define $\mathcal{B}\otimes\mathcal{C}$.
\begin{theorem}\label{tens-prod--Schur-kerns-is-SK}{\rm
In the notations and assumptions of Definition \ref{tens-prod--Schur-kerns},
$E_{\mathcal{B}; \ \cdot \ }\otimes E_{\mathcal{C}; \ \cdot \ }$ is an $(I,\mathcal{B}\otimes\mathcal{C})$--Schur kernel.
}\end{theorem}
\textbf{Proof}.
We have to prove that for any $n\in\mathbb{N}$ and any function
$$
X: h\in\{1, \dots,n\} \to X_{h}\in\mathcal{B}\otimes\mathcal{C}
$$
the complex valued kernel on $I\times \{1, \dots,n\}$ given by
\begin{equation}\label{Property-S-tens}
K_{(j,h),(i,k)}:=
E_{x;j,i}\left(X_{h}^*X_{k}\right)
\quad;\quad  (j,h), (i,k)\in I\times \{1, \dots,n\}
\end{equation}
is positive definite. Since elements of the form
$$
X_{k}=\sum_{m\in F}b_{k,m}\otimes c_{k,m}
$$
where $F$ is a finite set, are dense in $\mathcal{B}\otimes\mathcal{C}$, it is sufficient to
prove the statement for these elements.
For any function $z:(j,h)\in I\times \{1, \dots,n\} \to z_{j,h} \in\mathbb{C}$, one has
$$
\sum_{(j,h),(i,k)\in I\times \{1, \dots,n\}}\bar{z}_{j,h}K_{(j,h),(i,k)}z_{i,k}
=\sum_{(j,h),(i,k)\in I\times \{1, \dots,n\}}
\bar{z}_{j,h}E_{j,i}\left(X_{x,h}^*X_{x,k}\right)z_{i,k}
$$
$$
=\sum_{(j,h),(i,k)\in I\times \{1, \dots,n\}}
\bar{z}_{j,h}E_{j,i}\left(\left(\sum_{m'\in F}b^*_{h,m'}\otimes c^*_{h,m'}\right)
\left(\sum_{m\in F}b_{k,m}\otimes c_{k,m}\right)\right)z_{i,k}
$$
$$
=\sum_{(j,h),(i,k)\in I\times \{1, \dots,n\}}\sum_{m',m\in F}
\bar{z}_{j,h}E_{j,i}\left(\left(b^*_{h,m'}b_{h,m}\otimes c^*_{k,m'}c_{k,m}\right)\right)z_{i,k}
$$
\begin{equation}\label{Pos-S-tens1}
=\sum_{(j,h),(i,k)\in I\times \{1, \dots,n\}}\sum_{m',m\in F}
\bar{z}_{j,h}E_{\mathcal{B}; j,i}\left(b^*_{h,m'}b_{k,m}\right)
E_{\mathcal{C}; j,i}\left(c^*_{h,m'}c_{k,m}\right)z_{i,k}.
\end{equation}
Now define the new function
$$
(i,k,m)\in I\times \{1, \dots,n\}\times F \mapsto z_{i,k,m} := z_{i,k}.
$$
Then \eqref{Pos-S-tens1} becomes
\begin{equation}\label{Pos-S-tens2}
\sum_{j,i\in I }
\sum_{(h,m'),(k,m)\in \{1, \dots,n\}\times F}
\bar{z}_{j,h,m'}E_{\mathcal{B}; j,i}\left(b^*_{h,m'}b_{k,m}\right)
E_{\mathcal{C}; j,i}\left(c^*_{h,m'}c_{k,m}\right)z_{i,k,m}.
\end{equation}
Since
$$
|\{1, \dots,n\}\times F| = n\times |F|
$$
one can re--number the pairs $(k,m)\in \{1, \dots,n\}\times F$ so that they become in
one--to--one correspondence with the set $\{1, \dots,n|F|\}$.
With this numeration, \eqref{Pos-S-tens2} becomes
\begin{equation}\label{Pos-S-tens3}
\sum_{(j,h'),(i,k')\in I\times \{1, \dots,n|F|\}}
\bar{z}_{j,h'}E_{\mathcal{B}; j,i}\left(b^*_{h'}b_{k'}\right)
E_{\mathcal{C}; j,i}\left(c^*_{h'}c_{k'}\right)z_{i,k'}
\ge 0
\end{equation}
because, since $n$ in \eqref{Property-S} is arbitrary, both kernels
$(E_{\mathcal{B}; j,i}\left(b^*_{h'}b_{k'}\right))$ and
$(E_{\mathcal{C}; j,i}\left(c^*_{h'}c_{k'}\right))$ are positive definite, hence their Schur
product is by Schur lemma$.\qquad \square$\\

\section{The structure of superposition states}\label{sec_str_sup_state}
To each $x\in V$, we associate:\\
-- a finite set $D_{x}$ of \textbf{non--zero} vectors in $\mathcal{H}_{x}$,\\
-- a finite set $I$ of cardinality $d_{I}$,\\
-- a map
$$
h_{x, ( \ \cdot \ )}: i\in I \to h_{x, i}\in D_{x}
$$
and, for $\Lambda\subset_{fin} V$, define the vector
\begin{equation}\label{df-Psi-Lambda-gen}
\Psi_{\Lambda} :=\sum_{i\in I} \bigotimes_{x\in \Lambda} h_{x, i}
\in\bigotimes_{x\in \Lambda}\mathcal{H}_{x}=:\mathcal{H}_{\Lambda}
\end{equation}
and the associated positive linear functional on $ \mathcal{B}_{\Lambda}$ given by
\begin{equation}\label{df-psi-gen}
\psi_{\Lambda} = \langle \Psi_{\Lambda} , \ ( \ \cdot \ )\Psi_{\Lambda}\rangle
= \hbox{Tr}_{\mathcal{H}_{\Lambda}}(\Psi_{\Lambda}\Psi_{\Lambda}^* \ ( \ \cdot \ )).
\end{equation}
\begin{lemma}{\rm In the notations above
\begin{equation}\label{struc-PsiPsi*}
\Psi_{\Lambda}\Psi_{\Lambda}^* := \sum_{i,j\in I} \bigotimes_{x\in \Lambda} h_{x, i}h_{x, j}^*
\end{equation}
and, for
$$
\Lambda_{1}\supset \overline{\Lambda}
$$
$$
b_{\Lambda}\equiv\left(\bigotimes_{x\in \Lambda}b_{x}\right)\otimes 1_{\Lambda_{1}\setminus\Lambda}
\in \mathcal{B}_{\Lambda}\equiv \mathcal{B}_{\Lambda}\otimes 1_{\Lambda_{1}\setminus\Lambda}
$$
one has
\begin{equation}\label{psi-Lambda(bLambda1-gen}
\psi_{\Lambda_{1}}(b_{\Lambda})
=\sum_{i,j\in I} \prod_{x\in \Lambda}
\hbox{Tr}_{\mathcal{H}_{x}}\left(h_{x, i}h_{x, j}^* b_{x}\right)
\prod_{y\in \Lambda_{1}\setminus\Lambda}
\hbox{Tr}_{\mathcal{H}_{y}}\left(h_{y, i}h_{y, j}^* \right).
\end{equation}
In particular
\begin{equation}\label{struc-superp-sts}
\psi_{\Lambda}(b_{\Lambda})
=\sum_{i,j\in I} \prod_{x\in \Lambda}
\hbox{Tr}_{\mathcal{H}_{x}}\left(h_{x, i}h_{x, j}^* b_{x}\right).
\end{equation}
}\end{lemma}
\textbf{Proof}.
\eqref{struc-PsiPsi*} follows from \eqref{df-Psi-Lambda-gen} by distributivity.
$$
\psi_{\Lambda_1}(b_{\Lambda})
=\hbox{Tr}_{\mathcal{H}_{\Lambda_1}}
\left(\Psi_{\Lambda_1}\Psi_{\Lambda_1}^*  \bigotimes_{x\in \Lambda}b_{x}\right)
=\sum_{i,j\in I} \hbox{Tr}_{\mathcal{H}_{\Lambda_1}}
\left( \prod_{x\in \Lambda_1}h_{x, i}h_{x, j}^*  \prod_{x\in \Lambda}b_{x}\right)
$$
$$
=\sum_{i,j\in I} \hbox{Tr}_{\mathcal{H}_{\Lambda}}
\left(\bigotimes_{x\in \Lambda}h_{x, i}h_{x, j}^*b_x\right)  \hbox{Tr}_{\mathcal{H}_{\Lambda_1\setminus \Lambda}}
\left( \bigotimes_{y\in \Lambda_1\setminus \Lambda}h_{y, i}h_{y, j}^*\right)
$$
$$
=\sum_{i,j\in I} \prod_{x\in \Lambda}
\hbox{Tr}_{\mathcal{H}_{x}}\left(h_{x, i}h_{x, j}^* b_{x}\right)\prod_{y\in \Lambda_1\setminus\Lambda}
\hbox{Tr}_{\mathcal{H}_{y}}\left(h_{y, i}h_{y, j}^*\right).
$$
which is \eqref{psi-Lambda(bLambda1-gen}. And \eqref{struc-superp-sts} follows
from \eqref{psi-Lambda(bLambda1-gen} replacing $\Lambda_{1}$ by $\Lambda$ in $\psi_{\Lambda_1}$.
$\qquad \square$\\
\begin{theorem}\label{thm-Prop-S}{\rm
The family of linear functionals
\begin{equation}\label{df-Exji(bx)}
E_{x;j,i} : b_{x}\in \mathcal{B}_{x}\to
\hbox{Tr}_{\mathcal{H}_{x}}\left(h_{x, i}h_{x, j}^* b_{x}\right)\in\mathbb{C}
\quad;\quad  x\in V \ , \  j,i\in I
\end{equation}
enjoys {\bf property (S)} (\ref{Property-S}). Equivalently, for each $x\in V $ the map $E_x: (j,i) \mapsto E_{x, i,j}$ is an $(I,\mathcal{B}_{x})$--Schur kernel.
}\end{theorem}
\textbf{Proof}. In the notation of Definition \ref{df-Schur-kern}, for any $n\in\mathbb{N}$ and any function
$z:(j,h)\in I\times \{1, \dots,n\} \to z_{j,h} \in\mathbb{C}$, one has
$$
\sum_{(j,h),(i,k)\in I\times \{1, \dots,n\}}\bar{z}_{j,h}K_{x;(j,h),(i,k)}z_{i,k}
=\sum_{(j,h),(i,k)\in I\times \{1, \dots,n\}}
\bar{z}_{j,h}E_{x;j,i}\left(b_{x,h}^*b_{x,k}\right)z_{i,k}
$$
$$
=\sum_{(j,h),(i,k)\in I\times \{1, \dots,n\}}
E_{x;j,i}\left(\bar{z}_{j,h}b_{x,h}^*z_{i,k}b_{x,k}\right)
$$
$$
=\sum_{(j,h),(i,k)\in I\times \{1, \dots,n\}}
\hbox{Tr}_{\mathcal{H}_{x}}\left(h_{x, i}h_{x, j}^* \bar{z}_{j,h}b_{x,h}^*z_{i,k}b_{x,k}\right)
$$
$$
=\sum_{(j,h),(i,k)\in I\times \{1, \dots,n\}}
\left\langle h_{x, j}, \bar{z}_{j,h}b_{x,h}^*z_{i,k}b_{x,k}h_{x, i}\right\rangle
$$
$$
=\sum_{(j,h),(i,k)\in I\times \{1, \dots,n\}}
\left\langle z_{j,h} b_{x,h}h_{x, j}, z_{i,k}b_{x,k}h_{x, i}\right\rangle
$$
$$
= \left\langle \sum_{(j,h)\in I\times  \{1, \dots,n\}}z_{j,h} b_{x,h}h_{x, j},
\sum_{(i,k)\in I\times \{1, \dots,n\}} z_{i,k}b_{x,k}h_{x, i}\right\rangle
$$
$$
=\left\|\sum_{(i,k)\in I\times \{1, \dots,n\}} z_{i,k}b_{x,k}h_{x, i}\right\|^2 \ge 0.
$$
$\qquad \square$
\begin{corollary}\label{struc-hatE-Lambda}{\rm
Let $I$ be a finite set and, in the notation \eqref{df-B-Lambda-Bx}, let be given, for each $x\in V$
a Schur kernel $E_{x; \ \cdot \ }$ associated to the pair $(I,\mathcal{B}_{x})$. Then, for each $\Lambda\subset_{fin} V$, the map
$$
\hat{E}_{\Lambda}: \mathcal{B}_{\Lambda} \to (M_{d_{I}}(\mathbb{C}), \diamond)
$$
associated to the tensor product kernel $\bigotimes_{x\in\Lambda}E_{x; \ \cdot \ }$
according to Theorem \ref{pos-mps-Schur-kern} is characterized by
\begin{equation}\label{hat-ELambda}
\hat{E}_{\Lambda}\left(\bigotimes_{x\in\Lambda} b_{x}\right)
=\diamond_{x\in \Lambda}\hat{E}_{x}(b_{x})
\qquad;\qquad b_{x}\in\mathcal{B}_{x} \ , \ x\in\Lambda
\end{equation}
where each $\hat{E}_{x}: b\in \mathcal{B}_{x}\mapsto (E_{x,i,j}(b))\in  M_{d_{I}}(\mathbb{C})$ is the map
associated to according to Theorem \ref{pos-mps-Schur-kern}.
}\end{corollary}
\textbf{Proof}.
From Theorem \ref{tens-prod--Schur-kerns-is-SK} it follows by induction that
$E_{\Lambda; \ \cdot \ }:=\bigotimes_{x\in\Lambda}E_{x; \ \cdot \ }$ is a Schur kernel
and that the associated map
$\hat{E}_{\Lambda}:\mathcal{B}_{\Lambda} \to M_{d_{I}}(\mathbb{C})$ has the form
\eqref{hat-ELambda}.
$\qquad \square$
\begin{corollary}{\rm
In the identification of $M_{d_{I}}(\mathbb{C})$ with $\mathcal{B}(\mathbb{C}^{d_{I}})$
obtained by fixing a linear basis $e_{I}\equiv (e_{I,j})_{j\in I}$, for each $\Lambda\subset_{fin} V$, the state $\psi_{\Lambda}$ defined by \eqref{df-psi-gen} has the form
\begin{equation}\label{Schur-express-psi}
\psi_{\Lambda}(b_{\Lambda})
= \langle\hat{e}_{I} , \left(\diamond_{x\in \Lambda}\hat{E}_{x}(b_{x})\right)\hat{e}_{I}\rangle
\quad;\quad b_{\Lambda}:=\bigotimes_{x\in\Lambda}b_{x} \ , \ b_{x}\in\mathcal{B}_{x}
\end{equation}
where $\hat{E}_{x}$ is as in Corollary \ref{struc-hatE-Lambda} with $E_{x}$ given by
\eqref{df-Exji(bx)} and
$$
\hat{e}_{I}:=\sum_{j\in I} e_{I,j}
$$
is the non--normalized maximally entangled vector in the $e_{I}$--basis of $\mathbb{C}^{d_{I}}$.
}\end{corollary}
\textbf{Proof}. In the above notations
$$
\langle\hat{e}_{I}, \left(\diamond_{x\in \Lambda}\hat{E}_{x}(b_{x})\right)\hat{e}_{I}\rangle
=\sum_{j,i\in I}\langle e_{I,j} ,
\left(\diamond_{x\in \Lambda}\hat{E}_{x}(b_{x})\right)e_{I,i}\rangle
$$
$$
=\sum_{j,i\in I}\prod_{x\in \Lambda}E_{x;j,i}(b_{x})
=\sum_{j,i\in I}\prod_{x\in \Lambda}
\hbox{Tr}_{\mathcal{H}_{x}}\left(h_{x, i}h_{x, j}^* b_{x}\right)
=\psi_{\Lambda}(b_{\Lambda}).
$$
$\qquad \square$

\noindent\textbf{Remark}.
Formula \eqref{Schur-express-psi} suggests a natural \textbf{extension} of the superposition states obtained  by replacing the Schur  product's kernel in \eqref{df-Exji(bx)} by arbitrary Schur kernels.

\section{The infinite volume limiting state}\label{sec-inf-vol-lim}

If $\Lambda_{\alpha}$ is an increasing net of finite sub--sets in $V$, we write
$$
\Lambda_{\alpha}\uparrow\uparrow V
$$
to mean that the net $(\Lambda_{\alpha})$ definitively absorbs every finite set in $V$.
This implies
$$
\bigcup_{\alpha}\Lambda_{\alpha} = V.
$$
The simplified notation
$$
\Lambda_{1}\uparrow\uparrow V
$$
means that $\Lambda_{1}$ denotes the generic element of an increasing net
$\Lambda_{\alpha}\uparrow\uparrow V$.
\begin{lemma}\label{suff-cond-exist-lim}{\rm
The limit
\begin{equation}\label{lim-psi-Lambda1-gen}
\lim_{\Lambda_{1}\uparrow\uparrow V}\psi_{\Lambda_{1}}(b_{\Lambda})
=\sum_{i,j\in I} \prod_{x\in \Lambda}
\hbox{Tr}_{\mathcal{H}_{x}}\left(h_{x, i}h_{x, j}^* b_{x}\right)
\prod_{x\in \Lambda_{1}\setminus\Lambda}
\hbox{Tr}_{\mathcal{H}_{x}}\left(h_{x, i}h_{x, j}^* \right)
\end{equation}
exists for all $b_{\Lambda}\in\mathcal{B}_{\Lambda}$ if, for each $i,j\in I$, the limit
\begin{equation}\label{lim-psi-Lambda1-gen-ij}
\lim_{\Lambda_{1}\uparrow\uparrow V}\prod_{x\in \Lambda_{1}\setminus\Lambda}
\hbox{Tr}_{\mathcal{H}_{x}}\left(h_{x, i}h_{x, j}^* \right)
\end{equation}
exists. The converse is true if, for each $x\in V$, the $(h_{x, i})_{i\in I}$ are
linearly independent \textbf{but not a linear basis of} $\mathcal{H}_{x}$.
}\end{lemma}
\textbf{Proof}.
The implication \eqref{lim-psi-Lambda1-gen-ij} $\Rightarrow$  \eqref{lim-psi-Lambda1-gen} is clear
because $I$ is a finite set.\\
Conversely, if the limit \eqref{lim-psi-Lambda1-gen} exists for all $b_{\Lambda}\in\mathcal{B}_{\Lambda}$, the assumption on the vectors $h_{x, i}$ implies that
one can fix an $\overline{x}\in \Lambda$ and a pair $\bar i, \bar j\in I$ and choose the $b_{x}$
so that
\begin{equation}\label{choice-bx}
\mathrm{Tr}_{\mathcal{H}_{x}}\left(h_{x, i}h_{x, j}^*
b_{x}\right)
=\begin{cases}
\delta_{i,\bar i}\delta_{j,\bar j}c_{ij},   if  x=\overline{x}, \\
\\
0,    if    x\ne\overline{x}
 \end{cases}
\end{equation}
with $c_{ij}\ne 0$. In fact, since the $(h_{x, i})_{i\in I}$ are not a linear basis of
$\mathcal{H}_{x}$, for each $x\in V$ there is a non--zero vector $h_{x}$ orthogonal to
all the $(h_{x, i})_{i\in I}$. Choosing $b_{x}:=h_{x}h_{x}^*$ for $x\ne\overline{x}$,
the second identity in \eqref{choice-bx} is satisfied.\\
To satisfy the first identity in \eqref{choice-bx} we use the linear independence of the
$h_{\overline{x}, k}$. This guarantees that each such vector has the form
$$
h_{\overline{x}, k} = \hat{h}_{\overline{x}, k} + h^{\perp}_{\overline{x}, k}
$$
where $\hat{h}_{\overline{x}, k}$ is a linear combination of the set
$(\hat{h}_{\overline{x}, j})_{j\in I\setminus \{k\}}$ and
$h^{\perp}_{\overline{x}, k}$ is orthogonal to the linear span of this set.
Therefore, with the choice
$b_{\overline{x}}:=h^{\perp}_{\overline{x}, \bar{j}}(h^{\perp}_{\overline{x}, \bar{i}})^*$,
the first identity in \eqref{choice-bx} is satisfied.

\begin{lemma}{\rm
If, for each $x\in V$, the $(h_{x, i})_{i\in I}$ are an ortho--normal set,
condition \eqref{lim-psi-Lambda1-gen-ij} is satisfied and the limit positive functional is a
state and has the form
\begin{equation}\label{lim-psi-Lambda1-gen-orth}
\lim_{\Lambda_{1}\uparrow\uparrow V}\psi_{\Lambda_{1}}(b_{\Lambda})
=\sum_{i\in I} \prod_{x\in \Lambda}
\hbox{Tr}_{\mathcal{H}_{x}}\left(h_{x, i}h_{x, i}^* b_{x}\right)
=: \hat{\psi}_{\Lambda}(b_{\Lambda}).
\end{equation}
The family $(\hat{\psi}_{\Lambda})_{\Lambda\subset_{fin} V}$ is projective.
}\end{lemma}
\textbf{Proof}.
The assumption implies that, for each $i, j\in I$, the right hand side
of \eqref{lim-psi-Lambda1-gen-ij} is equal to $\delta_{ij}$.
Moreover, for each  $i\in I$, $\hbox{Tr}_{\mathcal{H}_{x}}\left(h_{x, i}h_{x, i}^* ( \ \cdot \ )\right)$
is a state on $\mathcal{B}_{x}$. Hence $\hat{\psi}_{\Lambda}$ is a sum of homogeneous product states on
$\mathcal{B}_{\Lambda}$. This implies projectivity.
$\square$

\noindent Define, for $x\in V$, the \textbf{interaction matrix at} $x$ by
$$
H_{x,i,j} :=\log\left(\hbox{Tr}_{\mathcal{H}_{x}}\left(h_{x, i}h_{x, j}^* \right)\right)
\iff \hbox{Tr}_{\mathcal{H}_{x}}\left(h_{x, i}h_{x, j}^* \right) = e^{H_{x,i,j}}
\quad;\quad i,j\in I
$$
where $\log$ denotes the principal value logarithm. Then for any $i,j\in I$
\begin{equation}\label{cond-lim-psi-Lambda1-gen-ij}
\lim_{\Lambda_{1}\uparrow\uparrow V}\prod_{x\in \Lambda_{1}\setminus\Lambda}
\hbox{Tr}_{\mathcal{H}_{x}}\left(h_{x, i}h_{x, j}^* \right)
=\lim_{\Lambda_{1}\uparrow\uparrow V}
e^{\sum_{x\in \Lambda_{1}\setminus\Lambda} H_{x,i,j} }
\end{equation}
in the sense that each limit exists iff the other one does and the equality holds.
\begin{lemma}\label{lm-constr}{\rm
Suppose that, for each $i,j\in I$, the limit \eqref{cond-lim-psi-Lambda1-gen-ij} exists
for each $\Lambda\subseteq_{fin} V$ and $\Lambda_{0}\subseteq\Lambda$ define:
\begin{equation}\label{df-bound-cond1}
\beta_{\Lambda^c;i,j}:=
\lim_{\Lambda_{1}\uparrow\uparrow V}\prod_{x\in \Lambda_{1}\setminus\Lambda}
\hbox{Tr}_{\mathcal{H}_{x}}\left(h_{x, i}h_{x, j}^* \right)
=\lim_{\Lambda_{1}\uparrow\uparrow V}
e^{\sum_{x\in \Lambda_{1}\setminus\Lambda} H_{x,i,j} };
\end{equation}
\begin{equation}\label{df-trans-bound-cond1}
\beta_{\Lambda,\Lambda_{0};i,j}:=
\prod_{x\in \Lambda\setminus\Lambda_{0}}
\hbox{Tr}_{\mathcal{H}_{x}}\left(h_{x, i}h_{x, j}^* \right)
=e^{\sum_{x\in \Lambda\setminus\Lambda_{0}} H_{x,i,j} }.
\end{equation}
Then the matrices $(\beta_{\Lambda^c;i,j})$ and $(\beta_{\Lambda,\Lambda_{0};i,j})$
are positive definite and satisfy
\begin{equation}\label{eq-trans-bound-cond1}
\beta_{\Lambda^c;i,j}\beta_{\Lambda,\Lambda_{0};i,j}
=\beta_{\Lambda_{0}^c;i,j}.
\end{equation}
Moreover, the limit \eqref{lim-psi-Lambda1-gen-ij} exists and is equal to
\begin{equation}\label{beta-psi-Lambda1-gen}
\hat{\psi}_{\Lambda}(b_{\Lambda})
:=\sum_{i,j\in I} \prod_{x\in \Lambda}
\hbox{Tr}_{\mathcal{H}_{x}}\left(h_{x, i}h_{x, j}^* b_{x}\right)
\beta_{\Lambda;i,j}
\end{equation}
and the family $(\hat{\psi}_{\Lambda})$ is projective.\\
\noindent Conversely let, for each $\Lambda\subseteq_{fin} V$ be given a positive definite matrix
$(\beta_{\Lambda^c;i,j})$ which satisfies \eqref{eq-trans-bound-cond1} with
$(\beta_{\Lambda,\Lambda_{0};i,j})$ given by \eqref{df-trans-bound-cond1}.
Then, for each $\Lambda\subseteq_{fin} V$, the linear functional \eqref{beta-psi-Lambda1-gen}
is positive and the family $(\hat{\psi}_{\Lambda})$ is projective.
}\end{lemma}
\textbf{Proof}.
If the limit \eqref{cond-lim-psi-Lambda1-gen-ij} exists, the fact that the matrices $(\beta_{\Lambda;i,j})$ and $(\beta_{\Lambda,\Lambda_{0};i,j})$ are positive definite follows
from Schur lemma. Equation \eqref{eq-trans-bound-cond1} follows from \eqref{df-bound-cond1}
and \eqref{df-trans-bound-cond1}. If $\Lambda_{0}\subseteq\Lambda$ and
$b_{\Lambda}=b_{\Lambda_{0}}\otimes 1_{\Lambda\setminus\Lambda_{0}}$, then
\eqref{df-bound-cond1} and \eqref{df-trans-bound-cond1} imply
$$
\hat{\psi}_{\Lambda}(b_{\Lambda_{0}}\otimes 1_{\Lambda\setminus\Lambda_{0}})
=\sum_{i,j\in I} \prod_{x\in \Lambda}
\hbox{Tr}_{\mathcal{H}_{x}}\left(h_{x, i}h_{x, j}^* b_{x}\right)
\beta_{\Lambda^c;i,j}
$$
$$
=\sum_{i,j\in I} \prod_{x\in \Lambda_{0}}
\hbox{Tr}_{\mathcal{H}_{x}}\left(h_{x, i}h_{x, j}^* b_{x}\right)
\prod_{x\in \Lambda\setminus\Lambda_{0}}
\hbox{Tr}_{\mathcal{H}_{x}}\left(h_{x, i}h_{x, j}^* \right)
\beta_{\Lambda^c;i,j}
$$
$$
=\sum_{i,j\in I} \prod_{x\in \Lambda_{0}}
\hbox{Tr}_{\mathcal{H}_{x}}\left(h_{x, i}h_{x, j}^* \right)
\beta_{\Lambda,\Lambda_{0};i,j}\beta_{\Lambda^c;i,j}
$$
$$
=\sum_{i,j\in I} \prod_{x\in \Lambda_{0}}
\hbox{Tr}_{\mathcal{H}_{x}}\left(h_{x, i}h_{x, j}^* \right)
\beta_{\Lambda_{0}^c;i,j}
=\hat{\psi}_{\Lambda_{0}}(b_{\Lambda_{0}}).
$$
Conversely, if $(\beta_{\Lambda^c;i,j})$ is  a positive definite matrix, then by Schur lemma
$\hat{\psi}_{\Lambda}(b_{\Lambda})$, defined by the right hand side of \eqref{beta-psi-Lambda1-gen},
is a positive linear functional and the projectivity of the family $(\hat{\psi}_{\Lambda})$
is proved as in the first part of the theorem.

\noindent\textbf{Remark}.
Lemma \ref{lm-constr} suggests a way to construct families of vectors
$(\Psi_{\Lambda})_{\Lambda\subset_{fin} V}$ which, in the limit $\Lambda\uparrow\uparrow V$,
produce states on the algebra $\mathcal{A}=\bigotimes_{x\in V}\mathcal{B}_{x}$.
In the following we discuss a way to realize this construction.
\begin{lemma}{\rm
Given any positive definite matrix $T :=(T_{i,j})_{i,j\in I}$, if $h= (h_{i,j})_{i,j\in I}$ is any
\textbf{right square root} of $T$, in the sense that
\begin{equation}\label{df-rght-sqrt}
hh^*=T
\end{equation}
the column vectors of $h$
\begin{equation}\label{df-col-vects-h}
h_{i} := (h_{i,m})_{m}  \qquad;\qquad i,m\in I
\end{equation}
satisfy the relation
\begin{equation}\label{<hj,hi>=Tij}
\langle h_{j} , h_{i}\rangle = T_{i,j}   \qquad;\qquad\forall i,j\in I.
\end{equation}
}\end{lemma}
\textbf{Proof}.
By assumption $T$ is positive definite. Let $h= (h_{i,j})$ be any right square root of $T$.
Define the column vectors of $h$ as in \eqref{df-col-vects-h}.
Then
$$
\langle h_{j} , h_{i}\rangle
=\sum_{h}\overline{h_{j,h}} h_{i,h}
=\sum_{h}  h_{i,h}(h^*)_{h,j}
=(hh^*)_{i,j}
=T_{i,j}.
\square
$$
\begin{lemma}\label{lm-constr-rght-sqrt}{\rm
For any Hermitean positive definite matrix $T=(T_{i,j})_{i,j\in I}$ on $\mathbb{C}^d$, there exists
an Hermitean matrix $H=(H_{i,j})$ on $\mathbb{C}^d$ such that $T$ has the form
\begin{equation}\label{H:=logT}
T =e^{H} = e^{U^*D_{H}U}
\end{equation}
where $U$ is a unitary matrix and
\begin{equation}\label{df-DH}
D_{H} = \hbox{diag}(h^{0}_{1}, \dots, h^{0}_{d})
\quad;\quad h^{0}_{j}\in\mathbb{R} \ , \ j\in \{1,\dots, d\}.
\end{equation}
Moreover, for any isometry $W$ on $\mathbb{C}^d$, the matrix
\begin{equation}\label{df-hW}
h:= e^{H/2}W^*
\end{equation}
is a right square root of $T$ and the entries of $H$ satisfy the inequality
\begin{equation}\label{|Hij|-le-Tr|DH|}
|H_{i,j}|\le \hbox{Tr}(|D_{H}|)
\qquad;\qquad\forall i,j\in I
\end{equation}
where
$$
|D_{H}| := \hbox{diag}(|h^{0}_{1}|, \dots, |h^{0}_{d}|).
$$
}\end{lemma}
\textbf{Proof}.
Since $T$ is Hermitean positive definite, by the spectral theorem it has the form
$$
T = U^*\hbox{diag}(t_{1}, \dots, t_{d})U
\quad;\quad t_{j}\in\mathbb{R}_+ \ , \ j\in \{1,\dots, d\}
$$
where $U$ is a unitary matrix. Defining
$$
D_{H} := \hbox{diag}(h^{0}_{1}, \dots, h^{0}_{d}):= \hbox{diag}(\log(t_{1}), \dots,\log(t_{d})).
$$
\begin{equation}\label{struc-Herm-matr}
H := U^*D_{H}U.
\end{equation}
One sees that \eqref{H:=logT} and \eqref{df-DH} hold.
And that $h$, defined by \eqref{df-hW}, is a right square root of $T$ follows from
$$
hh^*:= e^{H/2}W^* We^{H/2}  = e^{H} = T.
$$
Finally \eqref{H:=logT} implies that, for any $i,j\in I$.
$$
H_{i,j}
=\sum_{m,n} (U^*)_{i,m}h^{0}_{m}\delta_{m,n}U_{n,j}
=\sum_{m}\overline{U_{m,i}} h^{0}_{m}U_{m,j}.
$$
Therefore
$$
|H_{i,j}|
\le  \sum_{m}|\overline{U_{m,i}}| \  |h^{0}_{m}| \ |U_{m,j}|
\le  \sum_{m}  |h^{0}_{m}| =  \hbox{Tr}(|D_{H}|).
$$
which is \eqref{|Hij|-le-Tr|DH|}.
\begin{theorem}\label{thm_limit_state}{\rm
For each $x\in V$ fix:\\
-- the dimension of the fiber independent of $x$
$$
d:=\hbox{dim}(\mathcal{H}_{x}),
$$
-- a set $I$ such that
$$
d=|I| \quad;\quad I :=\{1,\dots,d\},
$$
-- a diagonal $d\times d$ matrix with real entries $D_{H_{x}}$,\\
-- a unitary $d\times d$ matrix  $U_{x}$.\\
So that the family of matrices $(D_{H_{x}})_{x\in V}$ satisfies the additional condition
\begin{equation}\label{constr-suff-cond0}
\sum_{x\in V}\hbox{Tr}(|D_{H_{x}}|) < +\infty.
\end{equation}
Define
$$
H_{x} := U_{x}^*D_{H_{x}}U_{x}
$$
and fix a right square root $h_{x}$ of $e^{H_{x}} = U_{x}^*e^{D_{H_{x}}}U_{x} $ as described in Lemma \eqref{lm-constr-rght-sqrt}.
If $(h_{x;i,m})_{m\in I}$ is the $i$--th  column vectors of $h_{x}$ ($i\in I$),
define the vector
\begin{equation}\label{constr-vects-h(xi)}
h_{x;i} := \sum_{m\in I} h_{x;i,m}e_{x;m}
\end{equation}
where $(e_{x;m})_{m\in I}$ is an ONB of $\mathcal{H}_{x}$. Then the family of vectors
associated to the vectors \eqref{constr-vects-h(xi)} as in \eqref{df-Psi-Lambda-gen}, i.e.
\begin{equation}\label{constr-Psi-Lambda-gen}
\Psi_{\Lambda} :=\sum_{i\in I} \bigotimes_{x\in \Lambda} h_{x, i}
\in\bigotimes_{x\in \Lambda}\mathcal{H}_{x}=:\mathcal{H}_{\Lambda}
\quad;\quad \Lambda\subset_{fin} V
\end{equation}
is such that the limit
\begin{equation}\label{constr-lim-psi-Lambda1-gen-ij}
\hat{\psi}(b_{\Lambda})
:=\lim_{\Lambda_{1}\uparrow\uparrow V}
\left\langle \Psi_{\Lambda_1},b_{\Lambda}\Psi_{\Lambda_1}\right\rangle
\end{equation}
exists for all $b_{\Lambda}\in\mathcal{B}_{\Lambda}$ and the family
$(\hat{\psi}_{\Lambda}(b_{\Lambda}))_{\Lambda\subset_{fin} V}$ is projective.
}\end{theorem}
\textbf{Proof}.
We know from Lemma \ref{suff-cond-exist-lim} that a sufficient condition for the existence of
the limit \eqref{constr-lim-psi-Lambda1-gen-ij} is that, for each $i,j\in I$, the limit
\begin{equation}\label{constr-suff-cond-lim}
\lim_{\Lambda_{1}\uparrow\uparrow V}\prod_{x\in \Lambda_{1}\setminus\Lambda}
\hbox{Tr}_{\mathcal{H}_{x}}\left(h_{x, i}h_{x, j}^* \right)
=\lim_{\Lambda_{1}\uparrow\uparrow V}
e^{\sum_{x\in \Lambda_{1}\setminus\Lambda} H_{x,i,j} }
\end{equation}
exists. From \eqref{|Hij|-le-Tr|DH|} and \eqref{constr-suff-cond0}, we know that
for all $i,j\in I$
$$
\sum_{x\in \Lambda_{1}\setminus\Lambda} |H_{x,i,j} |
\le \sum_{x\in V} |H_{x,i,j}|
\le \sum_{x\in V} \hbox{Tr}(|D_{H_{x}}|)
< +\infty.
$$
Therefore the series $\sum_{x\in \Lambda_{1}\setminus\Lambda} H_{x,i,j}$ is absolutely
convergent. Therefore for each $i,j\in I$, the limit \eqref{constr-suff-cond-lim} exists.
The projectivity of the family
$(\hat{\psi}_{\Lambda}(b_{\Lambda}))_{\Lambda\subset_{fin} V}$ follows from the convergence
of the series $\sum_{x\in \Lambda_{1}\setminus\Lambda} H_{x,i,j}$ with the same argument used in
Lemma \ref{lm-constr}.

\section{Homogeneous superposition states }\label{sec-Homog-case}

\textbf{Homogeneous bundles} on $V$ are characterized by the fact that there is a Hilbert space
$\mathcal{H}$ and for each $x\in V$, a unitary isomorphism
$$
J_{x}:\mathcal{H}_{x}\to \mathcal{H}
$$
a finite set $I_{0}$ and a set of vectors $(h_{j})_{j\in I_{0}}$ in $\mathcal{H}$ such that
$$
h_{x, j}= J_{x}^{*}h_{j} \qquad;\qquad\forall j\in I_{0} \ , \ \forall x\in V.
$$
In this case \eqref{psi-Lambda(bLambda1-gen} becomes
$$
\hbox{Tr}_{\mathcal{H}_{x}}\left(h_{x, i}h_{x, j}^* \right)
=\hbox{Tr}_{\mathcal{H}_{x}}\left(h_{x, i}h_{x, j}^* \right)
=\hbox{Tr}_{\mathcal{H}_{x}}\left(J_{x}h_{i}h_{j}^*J_{x}^{*} \right)
$$
\begin{equation}\label{df-beta-ij-hom}
=\hbox{Tr}_{\mathcal{H}}\left(h_{i}h_{j}^*\right)
=\langle h_{j},h_{i}\rangle
=:\beta_{ij}.
\end{equation}
Therefore
\begin{equation}\label{psi-Lambda(bLambda1-hom}
\psi_{\Lambda_{1}}(b_{\Lambda})
=\sum_{i,j\in I_{0}} \prod_{x\in \Lambda}
\hbox{Tr}_{\mathcal{H}_{x}}\left(h_{x, i}h_{x, j}^* b_{x}\right)
\beta_{ij}^{|\Lambda_{1}\setminus\Lambda|}.
\end{equation}
\noindent\textbf{Remark}.
The following Lemma shows that the condition that the $\beta_{ij}$ in
\eqref{df-beta-ij-hom} are constant in $(i,j)$, leads to non--interesting states.
\begin{lemma}\label{prod-functls1}{\rm
Let $\mathcal{K}$ be a Hilbert space,
$n\in \mathbb{N}$ and $c\in \mathbb{R}$ with $c>0$.
Then two vectors $h_{1}, h_{2}\in \mathcal{H}$ satisfy
\begin{equation}\label{cond-const-sc-prod}
\langle h_{j},h_{i}\rangle = c \qquad;\qquad\forall i,j\in \{1,2\}
\end{equation}
then they are proportional. \\
In particular, if the $\beta_{ij}$ in \eqref{df-beta-ij-hom} are
independent of $(i,j)$, the state $\psi_{\Lambda_{1}}$ in \eqref{psi-Lambda(bLambda1-hom} is
a product linear functional.
}\end{lemma}
\textbf{Proof}.
If the vectors $ h_{1}, h_{2}$ satisfy \eqref{cond-const-sc-prod}, then
\begin{equation}\label{constr-vects-const-sc-pr}
\langle h_{1},h_{2}\rangle^2 = c^2=  \langle h_{1},h_{1}\rangle\langle h_{2},h_{2}\rangle
\end{equation}
and by the Schwartz inequality they are proportional.\\
If the $\beta_{ij}=\beta$ in
\eqref{df-beta-ij-hom} are independent of $(i,j)$, then by the first part of the lemma, the
vectors $h_{i}$ are proportional to a single vector  $h\in\mathcal{H}$, hence for some
$t_{i}\in\mathbb{C}$, $h_{x, i}= t_{i}h_{x}=t_{i}J_{x}^{-1}h$. In this case
\eqref{psi-Lambda(bLambda1-hom} becomes
$$
\psi_{\Lambda_{1}}(b_{\Lambda})
=\sum_{i,j\in I_{0}}\overline{t_{j}}^{|\Lambda|} t_{i}^{|\Lambda|}
\beta^{|\Lambda_{1}\setminus\Lambda|}
\prod_{x\in \Lambda}
\hbox{Tr}_{\mathcal{H}_{x}}\left(h_{x}h_{x}^* b_{x}\right)
$$
which shows that $\psi_{\Lambda_{1}}$ is a product linear functional.\\
\noindent\textbf{Remark}.
Allowing a dependence of the $t_{i}$ on $x\in V$ only changes the homogeneous product linear
functional in the proof of Lemma \ref{prod-functls1} into a non-homogeneous one, but still of
product type.\\
\noindent\textbf{Remark}.
If we consider a weaker condition for which  (\ref{cond-const-sc-prod}) is satisfied only for $i\ne j$
then the vectors $(h_i)_{i\in I}$ are not necessary proportional. For the convenience of the reader, let  $(e_j)_{1\le j \le p}$ be an orthonormal family of vectors in $\mathcal{H}$. If $(\alpha_i)_{1\le i\le p}$ is the sequence of real number defined by
\begin{equation}
\alpha_1 =c ;\quad \alpha_{j} = c - (\alpha_1^2 + \alpha_2^2 + \cdots + \alpha_{j-1}^2);\quad j =2, \cdots, p-1.
\end{equation}
Then the  following vectors
$$
h_1 = e_1, \quad h_j =  \alpha_1e_1 +  \alpha_2 e_2 + \cdots +  \alpha_{j-1} e_{j-1} + e_j;\quad j =2, \cdots, p
$$
satisfy $<h_i, h_j> = c$ for  $i\ne j$. However, they are not proportional since they form  a linearly independent family in $\mathcal{H}$.
In this case the state in (\ref{psi-Lambda(bLambda1-hom})  becomes
\begin{equation}\label{psi_hom_hihj=c_i_ne_j}
\psi_{\Lambda_1}(a_\Lambda) = \sum_{i\in I} \prod_{x\in\Lambda}\hbox{Tr}(h_{x,i}h_{x,i}^{*}a_x )\beta_{ii}^{|\Lambda_1\setminus \Lambda|} +  \sum_{i\ne j} \prod_{x\in\Lambda}\hbox{Tr}(h_{x,i}h_{x,j}^{*}a_x )c^{|\Lambda_1\setminus \Lambda|},
\end{equation}
which is a combination of product  positive linear functionals.
\begin{theorem}\label{THM-prod-st1}{\rm
Let $x\in V\mapsto \mathcal{H}_{x}=J_{x}^*\mathcal{H}$ be an homogeneous bundle on $V$ and
let be given a finite set $I_{0}$ and a set of vectors $(h_{j})_{j\in I_{0}}$ in
$\mathcal{H}$. Define
\begin{equation}\label{h_x_copy_of_h}
h_{x, j}= J_{x}^{*}h_{j} \qquad;\qquad\forall j\in I_{0} \ , \ \forall x\in V.
\end{equation}
and suppose that the vectors $h_{j}$ satisfy the \textbf{generic condition}
\begin{equation}\label{gen-cond-beta}
i\ne j\Rightarrow
|\langle h_{j},h_{i}\rangle|=\beta_{ij} < \|h_{j}\| \ \|h_{i}\|
= \sqrt{\beta_{ii}}\sqrt{\beta_{jj}}.
\end{equation}
For $\Lambda\subset_{fin} V$, define the vector
\begin{equation}\label{df-Psi-Lambda-hom}
\Psi_{\Lambda} :=\sum_{i\in I} \bigotimes_{x\in \Lambda} h_{x, i}
\end{equation}
and the associated positive linear functional on $\mathcal{B}_{\Lambda}$ given by
\begin{equation}\label{df-psi-Lambda-hom}
\psi_{\Lambda} = \langle \Psi_{\Lambda} , \ ( \ \cdot \ )\Psi_{\Lambda}\rangle
= \hbox{Tr}_{\mathcal{H}_{\Lambda}}(\Psi_{\Lambda}\Psi_{\Lambda}^* \ ( \ \cdot \ )).
\end{equation}
Then, for each  $b_{\Lambda}:=\bigotimes_{x\in \Lambda}b_{x}\in \mathcal{B}_{\Lambda}$,
the limit
\begin{equation}\label{psi-Lambda(bLambda1-th}
\lim_{\Lambda_{n}\uparrow\uparrow V}
\frac{\psi_{\Lambda_{n}}(b_{\Lambda})}{\psi_{\Lambda_{n}}(1_{\Lambda_{n}})}
\end{equation}
exists, is independent of the sequence $(\Lambda_{n})$
with the above property and is equal to
\begin{equation}\label{lim-psi-Lambda1-th}
\frac{1}{\beta^{\Lambda}\sum_{i\in I_{0}, \beta_{ii}=\beta } 1}
\sum_{i\in I_{0}, \beta_{ii}=\beta } \prod_{x\in \Lambda}
\hbox{Tr}_{\mathcal{H}_{x}}\left(h_{x, i}h_{x, i}^* b_{x}\right)
\end{equation}
for some $\beta >0$. In particular, it is a convex combination of product states.
}\end{theorem}
\textbf{Proof}.
From \eqref{psi-Lambda(bLambda1-hom} it is known that
\begin{equation}\label{psi-Lambda(bLambda1-th2}
\psi_{\Lambda_{1}}(b_{\Lambda})
=\sum_{i,j\in I_{0}} \prod_{x\in \Lambda}
\hbox{Tr}_{\mathcal{H}_{x}}\left(h_{x, i}h_{x, j}^* b_{x}\right)
\beta_{ij}^{|\Lambda_{1}\setminus\Lambda|}.
\end{equation}
Therefore
$$
\psi_{\Lambda_{1}}(1_{\Lambda_{1}})
=\sum_{i,j\in I_{0}} \beta_{ij}^{|\Lambda_{1}|}.
$$
Denote
\begin{equation}\label{beta}
 \beta := \max\{|\beta_{ij}| \ : \ i,j\in I_{0}\}= \max\{|\beta_{ii}| \ : \ i\in I_{0}\}
\end{equation}
where the second identity follows from the Schwartz inequality. Then
$$
\frac{\psi_{\Lambda_{1}}(b_{\Lambda})}{\psi_{\Lambda_{1}}(1_{\Lambda_{1}})}
=\frac{1}{\sum_{i,j\in I_{0}} \beta_{ij}^{|\Lambda_{1}|}}
\sum_{i,j\in I_{0}} \prod_{x\in \Lambda}
\hbox{Tr}_{\mathcal{H}_{x}}\left(h_{x, i}h_{x, j}^* b_{x}\right)
\beta_{ij}^{|\Lambda_{1}\setminus\Lambda|}
$$
$$
=\frac{1}{\beta^{\Lambda_{1}}\sum_{i,j\in I_{0}} (\beta_{ij}/\beta)^{|\Lambda_{1}|}}
\sum_{i,j\in I_{0}} \prod_{x\in \Lambda}
\hbox{Tr}_{\mathcal{H}_{x}}\left(h_{x, i}h_{x, j}^* b_{x}\right)
\beta_{ij}^{|\Lambda_{1}\setminus\Lambda|}
$$
$$
=\frac{1}{\beta^{\Lambda}\sum_{i,j\in I_{0}} (\beta_{ij}/\beta)^{|\Lambda_{1}|}}
\sum_{i,j\in I_{0}} \prod_{x\in \Lambda}
\hbox{Tr}_{\mathcal{H}_{x}}\left(h_{x, i}h_{x, j}^* b_{x}\right)
(\beta_{ij}/\beta)^{|\Lambda_{1}\setminus\Lambda|}.
$$
From \eqref{df-beta-ij-hom} and \eqref{gen-cond-beta} one knows that, for $i\ne j$,
$$
|\beta_{ij}| < \beta.
$$
Therefore
$$
\lim_{\Lambda_{1}\uparrow\uparrow V}
\frac{\psi_{\Lambda_{1}}(b_{\Lambda})}{\psi_{\Lambda_{1}}(1_{\Lambda_{1}})}
=\frac{1}{\beta^{\Lambda}\sum_{i\in I_{0}, \beta_{ii}=\beta } 1}
\sum_{i\in I_{0}, \beta_{ii}=\beta } \prod_{x\in \Lambda}
\hbox{Tr}_{\mathcal{H}_{x}}\left(h_{x, i}h_{x, i}^* b_{x}\right).
$$
This proves \eqref{lim-psi-Lambda1-th}.\\
\textbf{Remark}
Theorem \ref{thm_prod_homog_beta_arb} below  propose a similar result to \ref{THM-prod-st1} replacing the generic condition \eqref{gen-cond-beta} by the weaker assumption that the $\beta_{ij}$ are real numbers but under the assumptions of Lemma \ref{lm-constr}.
\begin{theorem}\label{thm_prod_homog_beta_arb}{\rm
Under the assumptions and the notations of Lemma  \ref{lm-constr} and if $\psi_{\Lambda}$ is given by (\ref{df-psi-Lambda-hom}).
For each  $b_{\Lambda}:=\bigotimes_{x\in \Lambda}b_{x}\in \mathcal{B}_{\Lambda}$,
the limit (\ref{psi-Lambda(bLambda1-th}) exists, is independent of the sequence $(\Lambda_{n})$ and is equal to
\begin{equation}\label{lim-psi-Lambda1-th}
\frac{1}{\beta^{\Lambda}\sum_{i,j\in I_{0}, \beta_{ij}=\beta } 1}
\sum_{i,j\in I_{0}, \beta_{ij}=\beta } \prod_{x\in \Lambda}
\hbox{Tr}_{\mathcal{H}_{x}}\left(h_{x, i}h_{x, j}^* b_{x}\right)
\end{equation}
where $\beta $ is given by (\ref{beta}).
}\end{theorem}
\textbf{Proof}.
Lemma \ref{lm-constr} guarantees the existence of the limit  (\ref{psi-Lambda(bLambda1-th}), the rest of the proof can be done similarly to the proof of Theorem \ref{THM-prod-st1}. $\square$

\section{Mixing property for superposition states}\label{Sec_mixing}
In this section, we consider the multidimensional integer lattice for which $V=\mathbb{Z}^\nu$. A finite Hilbert space $\mathcal{H}$ is fixed and for each finite bounded region $\Lambda$,  unitary isomorphisms
$$
J_{\Lambda}:\mathcal{H}^{\otimes |\Lambda|}\to \mathcal{H}_{\Lambda}.
$$
If $\Lambda$ and $ \Lambda^{'}$ are bounded regions w $|\Lambda| = |\Lambda^{'}|$, we have the following identification
$$
J_{\Lambda}^{-1}(J_{\Lambda}(a)) \equiv J_{\Lambda^{'}}^{-1}(J_{{\Lambda^{'}}}(a)); \quad a\in \mathcal{H}^{\otimes |\Lambda|}.
$$
For $z = (z_1, z_2, \cdots, z_\nu )\in\mathbb{Z}^{\nu}$ we denote
$$
|z| = |z_1| + |z_2| + \cdots + |z_{\nu}|.
$$
Denote for each $r>0$
$$
D_r = \left\{ z\in\mathbb Z^\nu : \quad |z|\le r \right\}.
$$
For each $\Lambda^{'} \subset_{\hbox{fin}} V$ and $t>0$, then notation $J_t(\Lambda^{'})$ means any embedding of $\Lambda^{'}$ in $D_{t}^c$  \, .i.e.
\begin{equation}\label{Jt(Lambda')}
J_t(\Lambda^{'})\subset D_{t}^{c} ; \quad |J_t(\Lambda^{'})| = |\Lambda^{'}|.
\end{equation}
The set of all embedding of  $\Lambda^{'}$ in $D_{t}^{c}$  is identified to the set of all injective maps from  $\Lambda^{'}$ into $D_{t}^{c}$. It will be denoted by $\mathcal{E}(\Lambda^{'}, D_t^c)$.
For each $\Lambda\subset_{\hbox{fin}} \mathbb{Z}^{\nu}$, since
$$
\beta_{\Lambda^c, i,j} = \prod_{x\in \Lambda^c}\hbox{Tr}(h_{x,i}h^{*}_{x,j})\in\mathbb{C}.
$$
For $a_{\Lambda} = \bigotimes_{x\in\Lambda}a_x\in\mathcal{B}(\mathcal{H}_{\Lambda})$
\begin{equation}\label{psia_labbda}
\psi(a_{\Lambda}) = \sum_{i,j\in I}\prod_{x\in\Lambda} \left(\hbox{Tr}(h_{x,i}h^{*}_{x,i}) \hbox{Tr}(h_{x,j}h^{*}_{x,j})\right) \beta_{\Lambda^c, i,j}
\end{equation}
the  linear functional  $\psi$ is positive  and we assume  furthermore  that it is normalized .i.e
\begin{equation}\label{psi(1) =1}
\sum_{i,j\in I} \beta_{V, i,j} = 1.
\end{equation}
\begin{definition}\label{Mix_prop_def}
A state $\varphi$ on $\mathcal{A}_V$ is said to enjoy \textit{ the mixing property} if for each bounded regions $\Lambda, \Lambda^{'} \subset V$ and $a_\Lambda\in\mathcal{B}_\Lambda,  b_{\Lambda^{'}}\in \mathcal{B}_{\Lambda^{'}}$ the following property holds:
\begin{equation}\label{mix_prop}
\forall \varepsilon >0, \,   \exists t>0,  \, J_t\in \mathcal{E}(\Lambda^{'}, D_t^c) \quad   \left|\varphi(a_\Lambda  J_t(b_{\Lambda^{'}})) -  \varphi(a_\Lambda)\varphi( J_t(b_{\Lambda^{'}} ) ) \right| \leq \varepsilon.
\end{equation}
\end{definition}
\begin{definition} Let $(\alpha_p)_p$ be a net of states on $\mathcal{B}(\mathcal{H}^{\otimes p})$.
A state $\varphi$ on $\mathcal{A}_V$ is said to enjoy the  \textit{$\alpha$-mixing property} if for each bounded regions $\Lambda, \Lambda^{'} \subset V$ and $a_\Lambda\in\mathcal{B}_\Lambda,  b_{\Lambda^{'}}\in \mathcal{B}_{\Lambda^{'}}$ the following property holds:
\begin{equation}\label{alpha_mix_prop}
\forall \varepsilon >0, \, \exists t>0, \, \forall J_t\in \mathcal{E}(\Lambda^{'}, D_t^c)  \quad   \left|\varphi(a_\Lambda)  J_{t}(b_{J_{\Lambda}^{-1}(\Lambda^{'}})) -  \varphi(a_\Lambda)\alpha_{|\Lambda^{'}|}( b_{\Lambda^{'}}  ) \right| \leq \varepsilon
\end{equation}
\end{definition}
\begin{theorem}\label{thm_mixed_prop} Under the notations and assumptions of Lemma \ref{lm-constr}. Assume that for each finite region $\Lambda^{'}\subset_{\hbox{fin}} V$ and $b_{\Lambda^{'}}= \bigotimes_{y\in \Lambda^{'}}b_y\in\mathcal{B}_{\Lambda^{'}}$
the limit
\begin{equation}\label{cd_lim_bn}
\lim_{t\to + \infty}\prod_{y\in \Lambda^{'}}\mathrm{Tr}(h_{y ,i}h^\ast_{y ,j}b_y) = :\alpha^{(|\Lambda^{'}|)}(J^{-1}_{\Lambda^{'}}(b_{\Lambda^{'}}))
\end{equation}
exists and the limit $\alpha^{(|\Lambda^{'}|)}(J^{-1}_{\Lambda^{'}}(b_{\Lambda^{'}}))$ is independent of $i,j\in I$ . Then  $\alpha_{|\Lambda^{'}|}$ is a state of  product type on the algebra $\mathcal{B}(\mathcal{H}^{|\Lambda|})$.
 The state $\hat{\psi}$ is given by (\ref{constr-lim-psi-Lambda1-gen-ij}), enjoys the $\alpha$--mixing property (\ref{alpha_mix_prop}) and the mixing property (\ref{mix_prop}).
\end{theorem}
\textbf{Proof}. The functional $\alpha^{(|\Lambda^{'}|)}$ is a point-wise limit of positive functional of product type. This make it positive functionals. Moreover, since the left hand side of (\ref{cd_lim_bn}) is independent of the one-to-one  embedding $J_{t}$ of $\mathcal{B}(\mathcal{H}^{|\Lambda^{'}|})\equiv\mathcal{B}(\mathcal{H})^{|\Lambda^{'}|}$ into  $\mathcal{B}(\mathcal{H_{J_t(\Lambda)}})$. And since $\alpha^{(|\Lambda^{'}|)}$ is independent of the indices $i,j$ then  for $b_{\Lambda^{'}}\ge 0$ one has
$$
\alpha^{(|\Lambda^{'}|)}(b_{\Lambda}^{'}) =\lim_{t \to \infty} \prod_{y\Lambda^{'}}\mathrm{Tr}(h_{J_t(y) ,i}b_y h^\ast_{J_t(y) ,i}) \ge 0
$$
and
$$
\alpha^{(|\Lambda^{'}|)}(I_{{\Lambda}^{'}}) =\lim_{t \to \infty} \prod_{y\Lambda^{'}}\mathrm{Tr}(h_{J_t(y) ,i} h^\ast_{J_t(y) ,i})=1
$$
whenever $\mathrm{Tr}(h_{y ,i} h^\ast_{y ,i}) =1$ for each $y\in V$ and $i\in I$. Thus $\alpha^{(|\Lambda^{'}|)}$ is a state on $\mathcal{B}(\mathcal{H})^{|\Lambda^{'}|}$.\\
One has
$$
\hat{\psi}(a_{\Lambda} J_t(b_{\Lambda^{'}})) = \sum_{i,j}\left(\prod_{x\in   \Lambda}\mathrm{Tr}(h_{x,i}h^\ast_{x,j}a_x)\right)
\left( \prod_{y\in\Lambda^{'}}\mathrm{Tr}(h_{J_t(y) ,i}b_y h^\ast_{J_t(y) ,j})\right)\beta_{(\Lambda\cup J_t(\Lambda^{'}))^c, i, j}
$$
and
$$
\psi(a_{\Lambda}) \psi( J_t(b_{\Lambda^{'}})) = \left( \sum_{i,j} \prod_{x\in   \Lambda}\mathrm{Tr}(h_{x,i}h^\ast_{x,j}a_x)
\beta_{\Lambda^c, i, j}\right)\left(\sum_{k, l} \prod_{y\in   \Lambda^{'}}\mathrm{Tr}(h_{J_t(y), k}b_{y}h^\ast_{J_t(y),l})
\beta_{(J_t(\Lambda^{'}))^c, k, l}\right)
$$
Since $\sum_{z\in\mathbb{Z}^d}|H_{z, i,j}|< +\infty$ then as $t \longrightarrow +\infty$ one gets
\begin{eqnarray*}
\left\{ \begin{array}{llll}
\beta_{(\Lambda\cup J_t(\Lambda^{'}))^c, i, j}   \longrightarrow   \beta_{ \Lambda^c, i,j }; \\
\\
\beta_{(J_t(\Lambda^{'}))^c, k, l}  \longrightarrow \beta_{V, k, l}.\\
\end{array}
\right.
\end{eqnarray*}
From (\ref{cd_lim_bn}) one gets
\begin{equation}\label{lim_psi(ab)}
\lim_{t\to\infty}\hat{\psi}(a_{\Lambda} J_t(b_{\Lambda^{'}})) = \sum_{i,j}\left(\prod_{x\in   \Lambda}\mathrm{Tr}(h_{x,i}h^\ast_{x,j}a_x)\right)
\beta_{\Lambda^c, i, j}\alpha^{(|\Lambda^{'}|)}(J^{-1}_{\Lambda^{'}}(b_{\Lambda^{'}}))
\end{equation}
$$
=\psi(a_{\Lambda})\alpha^{(|\Lambda^{'}|)}(J^{-1}_{\Lambda^{'}}(b_{\Lambda^{'}})).
$$
Therefore, the state  $\psi$ satisfies the $\alpha$-mixing property (\ref{alpha_mix_prop}).
On the other hand, on has
\begin{eqnarray*}
\lim_{t\to\infty}\psi(a_{\Lambda}) \psi( J_t(b_{\Lambda^{'}}))& =& \left( \sum_{i,j} \prod_{x\in   \Lambda}\mathrm{Tr}(h_{x,i}h^\ast_{x,j}a_x)
\beta_{\Lambda^c, i, j}\right)\left(\sum_{k,l}\beta_{V, k, l} \alpha^{(|\Lambda^{'}|)}(J^{-1}_{\Lambda^{'}}(b_{\Lambda^{'}}))\right)\\ \nonumber
&=&  \left( \sum_{i,j} \prod_{x\in   \Lambda}\mathrm{Tr}(h_{x,i}h^\ast_{x,j}a_x)
\beta_{\Lambda^c, i, j}\right)\left(\sum_{k,l}\beta_{V, k, l}\right)\alpha^{(|\Lambda^{'}|)}(J^{-1}_{\Lambda^{'}}(b_{\Lambda^{'}})).
\end{eqnarray*}
And since
$$
\sum_{k,l}\beta_{V, k, l}=\psi(1)=1.
$$
Then
\begin{equation}\label{lim_psi(b)}
\lim_{t\to\infty}\psi(a_{\Lambda}) \psi( J_t(b_{\Lambda^{'}})) = \psi(a_{\Lambda})\alpha^{(|\Lambda^{'}|)}(J^{-1}_{\Lambda^{'}}(b_{\Lambda^{'}})).
\end{equation}
From (\ref{lim_psi(ab)}) and (\ref{lim_psi(b)}) for each  $\varepsilon>0$
there exists $t\ge0$ such that for each $J_t\in\mathcal{E}(\Lambda^{'}, D_t^c)$
$$
\left|\psi(a_{\Lambda} J_t(b_{\Lambda^{'}})  -  \psi(a_{\Lambda})\alpha^{(|\Lambda^{'}|)}
(J^{-1}_{\Lambda^{'}}(b_{\Lambda^{'}}))\right|<\varepsilon/2
$$
and
$$
\left|\psi(a_{\Lambda}) \psi( J_t(b_{\Lambda^{'}}))  -  \psi(a_{\Lambda})\alpha^{(|\Lambda^{'}|)}(J^{-1}_{\Lambda^{'}}(b_{\Lambda^{'}}))\right|<\varepsilon/2.
$$
Then, for each $J_t\in\mathcal(\Lambda^{'}, D_t^c)$
$$
\left|\psi(a_{\Lambda} J_t(b_{\Lambda^{'}})  -  \psi(a_{\Lambda}) \psi( J_t(b_{\Lambda^{'}}))\right|< \varepsilon.
$$
Therefore the state $\psi$ satisfies the mixing property (\ref{mix_prop}). $\square$\\
\textbf{Remark}
If the right hand side in (\ref{cd_lim_bn}) depends in the indices $i,j$,  then it defines a state of product type on $\mathcal{B}(\mathcal{H}_x)$ denoted by $\alpha^{(|\Lambda^{'}|)}_{i,j}(J^{-1}_{\Lambda^{'}}(b_{\Lambda^{'}}))$. In this case,
$$
\hat{\psi}(a_{\Lambda} J_t(b_{\Lambda^{'}})) \longrightarrow \sum_{i,j} \prod_{x\in   \Lambda}\mathrm{Tr}(h_{x,i}h^\ast_{x,j}a_x)
\beta_{\Lambda^c, i, j}\alpha^{(|\Lambda^{'}|)}_{i,j}(J^{-1}_{\Lambda^{'}}(b_{\Lambda^{'}}))
$$
while
$$
\psi(a_{\Lambda}) \psi( J_t(b_{\Lambda^{'}})) \longrightarrow   \left( \sum_{i,j} \prod_{x\in   \Lambda}\mathrm{Tr}(h_{x,i}h^\ast_{x,j}a_x)
\beta_{\Lambda^c, i, j}\right)\left(\sum_{k,l}\beta_{V, k, l}\alpha_{k, l}^{(|\Lambda^{'}|)}(J^{-1}_{\Lambda^{'}}(b_{\Lambda^{'}}))\right).
$$
Therefore, the mixing properties (\ref{alpha_mix_prop}) and (\ref{mix_prop}) are not necessarily satisfied.

\end{document}